\documentclass[
  journal=largetwo,
  manuscript=article-type,
  year=2020,
  volume=37,
]{cup-journal}

\usepackage{amsmath}
\usepackage[nopatch]{microtype}
\usepackage{booktabs}

\title{Some Findings from the Longitudinal Migration of Starspots}

\author{N.\,\"{O}. Kaya}
\affiliation{Ege University, Science Faculty, Department of Astronomy and Space Sciences, 35100 Bornova, \.{I}zmir, Turkey.}
\email[N.\,\"{O}. Kaya]{nurhanozlemk@gmail.com}

\author{H. A. Dal}
\affiliation{Ege University, Science Faculty, Department of Astronomy and Space Sciences, 35100 Bornova, \.{I}zmir, Turkey.}


\begin{document}

\begin{abstract}
We present results regarding the longitudinal migrations of cool stellar spots that exhibit remarkable oscillations and explore their possible causes. We conducted analyses using high-quality data from nine target systems of various spectral types, spanning from F to M, which were observed by the Kepler Satellite. The systems in which the behaviour of the spots was examined are as follows: KIC 4357272, KIC 6025466, KIC 6058875, KIC 6962018, KIC 7798259, KIC 9210828, KIC 11706658, KIC 12599700, and KIC 8669092. Basic stellar parameters were calculated from light curve analysis using the PHOEBE V.0.32 software, and light curves were modelled to obtain sinusoidal variations occurring out-of-eclipses phases, induced by rotational modulation. Subsequently, we calculated the minimum times of the obtained sinusoidal variations using the Fourier transform. The distributions of ${\theta}_{min}$ corresponding to these minimum times over time were computed using linear fits to determine the longitudinal migrations of the spotted areas. We then compared the longitudinal migration periods with the stellar parameters found in the literature. In addition, we also found a secondary variation in the spot migrations apart from the linear models. Our results revealed that the longitudinal migration periods vary in relation to the $B-V$ colour index of the stars. 
\end{abstract} 

\section{Introduction}\label{S1}

Several eclipsing binaries have components exhibiting stellar spot activity, and these targets are known as RS CVn or BY Dra systems. The chromospheric activity was first found in these systems by \citet{kron52}, who demonstrated a sinusoidal-like variation at out-of-eclipse phases in the light curve of YY Gem. \citet{kron52} explained the sinusoidal-like variation at out-of-eclipse phases as a heterogeneous temperature distribution on the stellar surface, which is called 'BY Dra Syndrome' by \citet{Kun75}. BY Dra Syndrome caused by rotational modulation was confirmed by the later work of \citet{Tor73}, \citet{Bopp73}, \citet{Vogt75}, \citet{Frie75}.

Stars exhibiting chromospheric activity are typically cool stars ranging from spectral types F2 to M8, which rotate rapidly and often have convective outer layers. One of the most prominent features observed in stars with chromospheric activity is cool stellar spots. These spots also observed on the Sun, are ideal cases for determining the rotation periods of stars, such as differential rotation. Additionally, they play a significant role in understanding stellar dynamos \citep{Thomas08}. Detecting and modelling stellar spots on the surfaces of stars is a phenomenon of great importance for understanding the evolution of cool stars and comprehending various unknown parameters.

We analyzed nine chromospherically active targets identified as eclipsing binary systems by \citet{Wat06}. There is not much research available in the literature regarding the systems we examined. Additionally, many studies examining the data obtained from the $Kepler$ database have determined observational indicators related to flares and spot activity of some systems. For instance, 87 flares from the observation data of the active component of the KIC 7798259 were detected by \citet{Gao16}. \citet{Shi13} detected four flares from the observation data of KIC 4357272, while \citet{Gao16} found 58 flares from the same data. \citet{Balo15}, \citet{Gao16}, and \citet{Kaya19}, who analyzed the light variations of KIC 8669092, known as KOI 68AB, stated that the target is a binary system with an active chromospheric component. Also, \citet{Kaya19} detected 313 flares exhibited by the active component of KOI 68AB during an observation period of approximately 4 years. Furthermore, they observed the longitudinal and latitudinal movement of the two-spotted areas on the surface of the active component. In general, the physical parameters of the examined systems are listed in Table 1. Their $B-V$ values in the eighth column were calculated based on their $T_{eff}$ temperatures obtained from the MAST database using the calibrations provided by \citet{Tok00}, \citet{Cox00}.

In this study, photometric observation data from the $Kepler$ database for systems with different temperatures were examined. In Section 2.1, the light curves of the systems were analyzed, and basic physical parameters for the components were obtained. In Sections \ref{S2.2} and \ref{S2.3}, the longitudinal migration periods of spots on the active components of the systems were detected and calculated. Additionally, in Section \ref{S2.4}, frequency analysis was conducted to determine if there are any effects other than longitudinal ones. In Section \ref{S3}, the longitudinal migration periods obtained were compared with the basic parameters of the systems, namely, $B-V$, $log(g)$, and periods, and the study investigated whether there is a change according to a specific function.

\section{Data and Analyses}\label{S2}

The Kepler Mission was developed to discover exoplanets \citep{Koc10, Cal10}. The brightness values of nearly 150.000 targets were measured photometrically by the Kepler satellite. The importance of these photometric observations is that they have high-quality \citep{Jen10a, Jen10b} because the observational strategy of the Kepler satellite minuses aliasing in the period determination, which provides better frequency resolution that is important for the analyses in this study. Therefore, these data are essential for astrophysicists looking for exoplanets and astrophysicists studying different types of variable stars \citep{Sla11, Mat12}. In this study, we use the short-cadence (58,89 seconds) and the long-cadence (29,4 minutes) $Kepler$ photometry of nine targets available at the Kepler Eclipsing Binary Catalog. The data taken from the Kepler Eclipsing Binary Catalog \citep{Mat12, Kir16} were de-trended. The light curves created from the long-cadence data of each system are given in Figure 1, and the light curves were plotted by using the de-trended data as given in the Kepler Eclipsing Binary Archive. The data were arranged into different formats for the different analyses, as it is described in the following sections.

\subsection{Light Curve Analyses}\label{S2.1}

As seen in Figure 1, there are sinusoidal changes at out-of-eclipse phases observed in the light curve of nine binary systems examined. There may be three different reasons for these sinusoidal changes observed in the light curves of binary systems. The first of these is a reflection effect caused by the change of shape of the stars as they approach each other. However, we have eliminated this as the reflection is not maximum in the primary and secondary minima in the light curves of binary systems. Secondly, the sinusoidal changes due to the pulsation of stars can be observed. However, when the $log(g)$ and $T_{eff}$ values, which are the absolute parameters of the systems, are examined, it is seen that it does not localize with the instability strip in the $Hertzsprung-Russell$ diagram and localizes with the cold area where more active stars are present. Finally, there may be a change due to spot activity. Both their estimated positions in the $Hertzsprung-Russell$ diagram and the presence of flare activity observed in all of them increase the likelihood that the out-of-eclipse phases' sinusoidal change in the light curves of the systems may be caused by spot activity. Therefore, we proceeded with our analysis of the presence of spot activity in these systems.
 
 We concentrated on the variations seen at out-of-eclipse phases to obtain the rotational modulation effects due to the stellar spot activity. Among the targets we analysed in this study, four of nine systems have large eclipse depths, which seem to be enough to affect the shape of variations caused by the stellar spot activity. Because of this, we tried to model the light curves of the four systems by the light curve analysis. When analyzing the light curves of the targets (KIC 6962018, KIC 7798259, KIC 9210828, and KIC 6058875), the first step is to select consecutive cycles that are the least affected by rotational modulation. Then, we used the PHOEBE V.0.32 software \citep{Prs05}, which depends on the method used in the 2014 version of The Wilson-Devinney code \citep{Wil15}, to analyze the light curves. Initial attempts showed that astrophysically and statistically acceptable solutions for the four systems could only be obtained in Mode 2 (detached system). 

In the literature, there are different values for the temperatures of the primary components. Therefore, we used the temperatures obtained from \citet{Arm14} for each system. These temperature values were fixed as the temperatures of the primary components, while the temperatures of the secondary components were taken as adjustable free parameters. In addition to the $B-V$ values listed in Table 1, estimated mass values for primary and secondary components were calculated using calibrations by \citet{Tok00} and \citet{Cut03}, and their associated mass ratio ($q$) values were taken as adjustable free parameters in the analysis. Considering the possible spectral types of the components, the albedos ($A_{1}$ and $A_{2}$) and the gravity-darkening coefficients ($g_{1}$ and $g_{2}$) are taken for the stars with the convective envelopes \citep{Lucy67, Ruc69}, while the non-linear limb-darkening coefficients ($x_{1}$ and $x_{2}$) are taken from \citet{van93}. The dimensionless potential values for the components ($\Omega_{1}$ and $\Omega_{2}$), the fractional luminosity ($L_{1}$) of the primary components, and the inclination ($i$) of the systems are taken as the adjustable free parameters. As a result of the analysis, the obtained parameters are reported in Table 2, and the synthetic light curves with the observation data are shown in Figure 2.

We performed light curve analyses for KIC 6962018, KIC 7798259, KIC 9210828 and KIC 6058875 to obtain their variations out-of-eclipse phases, for which synthetic light curves were needed. In the case of KIC 8669092, one of the remaining five systems, the light curve analysis was recently done by \citet{Kaya19}. We used the synthetic curve they derived to obtain the variations seen at out-of-eclipse phases of this system. However, in the cases of the other four targets (KIC 12599700, KIC 11706658, KIC 4357272 and KIC 6025466), which have the minima with remarkable shallow amplitudes, we just removed the eclipsing minima parts from their light curves to obtain their variation seen at out-of-eclipse phases due to the stellar spot activity.

We would also like to add that the temperature values of the components calculated in the light curve analysis were obtained so that the light curve could be modelled properly and extracted from the data properly in the next stages. In the next stages, interpretation was made using the $T_{eff}$ temperatures of the systems and their corresponding $B-V$ values.

\subsection{Rotational Modulation and Stellar Spot Activity}\label{S2.2}

We used just the long cadence data for all targets to examine the variation at out-of-eclipse phases due to the stellar spot activity. The common feature of the targets' data is the rotational modulation effects seen at out-of-eclipses in their light variation apart from the eclipses. Considering the spectral types of the components together with their flare activities, the observed rotational modulation should have originated from the stellar spots, which are also the main issue of this study. Firstly, we removed the flare activity, which is seen as a sudden and rapid increase of the stellar flux in the light variation to reveal the rotational modulation effect. Most of the examined systems are presented as Algol-type binary systems in the SIMBAD database. However, the four targets (KIC 12599700, KIC 11706658, KIC 4357272 and KIC 6025466) have so shallow eclipse depths that the minima are nearly unnoticeable in the light curves, while the five targets (KIC 6962018, KIC 7798259, KIC 9210828, KIC 6058875 and KIC 8669092) have light curves with large eclipse depths. In order to obtain the variation at out-of-eclipse phases, we used the synthetic light curves derived by the light curve analyses to take out the residuals. On the other hand, in the case of the four targets with shallow eclipse depths, we just removed the data from the phase of 0.45 to 0.55 and from the phase of 0.95 to 0.05.

Considering the existence of the rotational modulation effect due to chromospheric activity, the sinus-like light variations at out-of-eclipse phases were modelled by using the Fourier transform. The expression defined by Equation (1) was used to model this variation \citep{Mor85}:

\begin{equation}
    L({\theta})=A_{0}+~\sum_{j=1}^{N} A_{j} cos(j{\theta}) +~\sum_{j=1}^{N} B_{j} sin(j{\theta})
	\label{Eq1}
\end{equation}

where $A_{0}$ is the zero point, $\theta$ is the phase, while $A_{j}$ and $B_{j}$ are the amplitude parameters. It is suggested that the $cos(\theta)$ term in this equation is related to the stellar spots \citep{Hall90}.  If there is only one-spotted area on the stellar surface, the $cos(\theta)$ term is expected to dominate. However, \citet{Hall90} suggests that the $cos(2\theta)$ term can be much more dominant in the case of two-spotted areas separated by $180^\circ$ from each other on the surface of the active component \citep{Dal12}.

\subsection{Detection of Longitudinal Spot Migration}\label{S2.3}

In the second step, we removed eclipse depths from observation data as described in Section 2.2. The observations that were scattered due to technical problems were removed from the data. Secondly, we tried to model the residual data using the Fourier method to obtain the best model. We recognized that the best models were generally derived by taking into account two-spotted areas on the stellar surface. Then, we obtained the synthetic Fourier curves for the residual data. In the last step, using the least-squares method, we determined the minima times for each target from the sinusoidal variation seen in the synthetic data and calculated the phases corresponding to their times according to their orbital period. A sample light variation for the sinusoidal variation for KIC 12599700 is given in Figure 3. In the figure, a cycle is given where eclipses have been removed and only the spotted-induced sinusoidal change is dominant. We clearly observe nested sinusoidal variations where two-spotted regions are evidently influential. Our purpose in the fitting we have made is to consider these two spotted areas separately and determine their actual longitudinal positions on the component's surface. Therefore, by examining two separate Fourier models created by two frequency values, we can obtain the longitudinal value for two separate spotted areas. The blended version of these two Fourier models represents the cycle we are analyzing and is represented in blue in the figure.

If someone examines the phases of the minima times of the sinusoidal variations, it will be seen that the minima times migrate from the phase of 1.0 to 0.0 with an overall descending trend, as described previously by \citet{Berd03}. In the case of the second active region, whose corrected migratory movement is seen in the lower-left panels, the minima phases of the sinusoidal variation could be migrated several times from phase 1.0 to 0.0, as can be seen in the upper panel of Figure 4. Because of this, we identified the cycle of the first migration from the phase of 1.0 to 0.0 as zero and 1 for the second migration, etc. However, the primary active region moves in a narrow band in the upper panel of the figure. An attempt was made to obtain a linear model by adding a cycle for each movement from the phase of 0.0 to 1.0. In this study, this process was named "${\theta}_{min}$". Thus, we obtained general trends of migration during these four years' observations. An example of the linear distributions and linear models is shown in the second top panel of Figure 4. Then we fitted these descending trends with the linear function. In the last step, we obtained the residuals by extracting the model fits from the migration trends.

In the literature, there are several studies, following a similar way to obtain the stellar spot migration \citep{Mak89, Tas01, Ros09, Bal15, Dal18, Yol21}.

\subsection{Second-order Changes Seen in the Spot Migration}\label{S2.4}

The residuals of ${\theta}_{min}$ were obtained by extracting the linear models from the ${\theta}_{min}$ variations, which are the sign of the longitudinal migration. Looking at the variation of ${\theta}_{min}$ residuals vs time, it is seen that there are systematic variations in the residuals. As seen in the third panel of Figure 4, regular sinusoidal variations in the ${\theta}_{min}$ residuals are noticeable. The variations observed in the ${\theta}_{min}$ residuals indicate that the ${\theta}_{min}$ values are shifted from linear trends over time, possibly due to latitudinal effects and changes in the position of the spots.

To understand the source of the variations seen in the ${\theta}_{min}$ residuals, the nature of the secondary variations must first be determined. For this reason, frequency analysis was conducted to determine whether the variations seen in the ${\theta}_{min}$ residuals are regular, or not. The analyses were done by using the Period04 software \citep{Lenz05}, which depends on the method of Discrete Fourier Transform (DFT; \citet{Sca82}). Iterations were continued until we were sure that the signal-to-noise ratio (SNR) remained at 3 or lower in the analyses performed with the program. As a result, in the frequency analysis of the ${\theta}_{min}$ residuals of each component, the frequency values that best fit the sinusoidal changes were found. The frequency and period values are reported in Tables 4-6. A sample for the models derived for the ${\theta}_{min}$ residuals by using these frequencies is shown in the bottom panel of Figure 4.

\section{Results and Discussion}\label{S3}

The longitudinal migrations of the stellar spots of the nine systems were examined, and it is shown in detail for KIC 6962018 as an example in Figure 4. To determine the spot migrations, the ${\theta}_{min}$ variations for each spot were obtained and given in the middle panels in the figure. Using the derived linear function, we calculated the migration periods of each spot ($P_{mig}$) and listed them in Table 3. Upon examination of the table, we observed that $P_{mig_1}$, $P_{mig_2}$ and $P_{mig_3}$ values were close in some systems, but there was almost a 2 times difference in others. Considering that such a difference is unlikely to occur on the same star, we classified the targets into three distinct groups.

As explained in Sections 2.3 and 2.4, we determined the two migration periods. As seen in Table 3, the migration periods of the spots detected on the active component in the KIC 4357272, KIC 6025466, KIC 6058875, KIC 8669092, and KIC 11706658 systems are very close to each other. Therefore, we consider these five systems as one group. When we consider the longitudinal migration periods of the spots we detected on the active components of the KIC 7798259 and KIC 9210828 systems, we have grouped these two systems as the second group due to the migration periods of the spots are not compatible with each other, and one of them is twice as high as the other for each system. In the case of KIC 7798259, the ${\theta}_{min}$ value of the first spot area completes three cycles in four years, while the second spot area completes six cycles. This may indicate that the first spot area completes three rotations around the star, while the second spot area completes six rotations. We think that these two spotted areas could not be on the same component since the stellar rotation period cannot differ twice at different latitudes. In the case of both KIC 7798259 and KIC 9210828, it can be observed that the temperatures of the primary and secondary components are similar. Therefore, there may be activity on both components of these systems.

The situation for KIC 6962018 and KIC 12599700, which we consider as the third group, is slightly different from the others. As seen in Figure 4, we detected two spotted areas in the case of KIC 6962018. These two spotted areas exhibit almost synchronous migration with a phase difference of about 0.50. The first spotted area should be located toward the upper latitudes on the active component, where it is less affected by differential rotation. However, the second spotted area shows a migration in a dramatic linear trend from 1.0 to 0.0, similar to other systems. This may mean that the second spotted area should be located toward the lower latitudes, where the differential rotation effect is dominant. However, for KIC 12599700, there is a different situation. In our results for this system, two spotted areas are seen in a phase interval of 0.40 at the beginning of the cycle. The first spotted area is located between phases 1.00 to 0.60, while the second is located between phases 0.60 to 0.40. Upon closer examination, it becomes evident that the second spotted area is not a single area on the component, but should be two different spotted areas that separate toward the end of the cycle. After the 500th day of the cycle, the second spotted area starts to exhibit two different longitudinal migrations in a phase interval of approximately 0.30. A sudden transition like this indicates a phase shift of the area or the formation of a new spotted area at a different active longitude. At this point, 3 different active longitudes are observed on the stellar surface. According to the two migration periods listed in Table 3, the migration period of the first spotted area is longer than the others. The migration periods of the second and third ones are closer to each other. According to a possible scenario, it can be interpreted that the first spotted area is located on one active component, while the second and third ones are located on the other active component.

Although the spot migrations exhibit dominantly linear variation, there are also secondary variations in a sinusoidal form. These variations are seen in the results of all targets. We tried to determine their possible frequencies from the ${\theta}_{min}$ residuals to understand whether these variations are systematic phenomena caused by any real physical conditions of the targets. The obtained frequencies are tabulated in Tables 4-6. When Tables 4-6 are examined carefully, the periods found for the sinusoidal variation indicate that the spotted areas are at the same component for each target. The periods of sinusoidal variation in the ${\theta}_{min}$ residuals are close to each other if the spotted areas are located on the same component. They are quite different from each other if the spotted areas are located at different components. Looking at Figure 4, certain sinusoidal changes in the ${\theta}_{min}$ residuals are remarkable. Considering that the longitudinal migration of the spots is not involved at this stage, it is thought that these sinusoidal changes may be caused by different latitudinal effects. Considering the studies of \citet{cole14} and \citet{Ozav18}, there are arguments supporting our thinking. Depending on the latitude where the spots are located, certain drift rates are seen in the longitudinal observation data. Therefore, it is thought that in observation data where the longitudinal effect is removed, certain changes caused by the latitudinal effects can be seen in the observation data. \citet{cole14} stated in their study that these effects could be: $(i)$ surface differential rotation, $(ii)$ successive emergence of multiple magnetic flux tubes (out of the same toroidal flux bundle) in the pro-grade direction, and $(iii)$ pro-grade azimuthal propagation of dynamo waves \citep{Ozav18}.

Using the obtained frequencies, we fitted the ${\theta}_{min}$ residuals. The sample is shown in the bottom panels of Figure 4. When each of them is examined one by one, it seems that the variations of the residuals have large amplitudes as much as 0.20 in phase values. At this point, we think that these large amplitudes can occur because the spots separate into different areas around the active longitudes, which can reveal themselves as a wide band. If a spot migrates in the wide band, we can see it as a sinusoidal variation in the ${\theta}_{min}$ residuals. The determined frequencies demonstrated the migration of a spot inside the active longitude band.

On the other hand, in the case of KIC 11706658, the amplitude of the ${\theta}_{min}$ residuals is remarkably large. The reason for this large amplitude cannot be the variation in the spot location within an active longitude band. It is most probable that the spotted areas changed their locations in latitude due to the differential rotation. The spotted areas located toward lower latitudes rotate with faster periods than the general rotation period of the target, which causes an increase in spot migration. However, the ${\theta}_{min}$ residuals do not only exhibit an increase but also a decrease. According to our understanding, supergranule-scaled convection currents must have dislocated the spotted areas toward upper latitudes or previous longitudes. \citet{Brad14} stated that spotted areas can be easily disintegrated due to supergranule-scaled convection currents.

The distributions of longitudinal migration periods versus the $B-V$ colour index, $log(g)$, and rotational periods of the targets were examined. The longitudinal migration periods are plotted vs the stellar $B-V$ colour index in the right column of Figure 5. Longitudinal migration periods between $0^m.50$ and $1^m.15$ stellar $B-V$ colour indexes appear to be shorter than 800 days. It is well known that the stars with a $B-V$ index higher than $1^m.50$ are completely convective cool stars. In the case of this type of target, the turbulence dynamo is dominant, and it is not clear how the differential rotation profile on their surfaces works.

In this study, one of the interesting targets is KIC 12599700. As seen from the right column of Figure 5, the migration periods found for the first and second spotted areas follow a rule as expected, according to the colour indexes of the components. Here, we expected that the migration periods follow a linear trend. However, the regression calculations revealed that the periods follow polynomial functions with secondary degrees, which are represented by red lines in Figure 5. On the other hand, the temperature of the primary component is reported to be 3758 K, and that of the secondary component is 3759 K \citep{Arm14}. However, the migration period of the third spotted area plotted as a star in the figure, is not where it should be on the theoretical curve. The target is a binary system whose components are almost identical, yet the third spotted area seems incompatible with the temperatures of these components. It should correspond to a component with a $B-V$ colour index of $1^m.80$. Therefore, we focused on the light curve of the target. The depths of the primary and secondary eclipses are almost equal but very shallow. Considering the temperatures of the components, one possible explanation for these shallow eclipse depths is that the system has an unseen third component. If there is an unseen third component in the system, it would be expected to have little light excess in the total light, as is often seen in such astrophysical situations. In this case, the third body must be a very cool M dwarf. This scenario might explain the star in the figure. If this is considered to be the case, these period changes will appear closer to a linear distribution versus the $B-V$ colour indexes.

In the left panels of Figure 5, we compare the two migration periods to the stellar $log(g)$ values. As seen in the middle left panel, the stellar spot migration periods generally increased with increasing $log(g)$ values. The migration period values are widely scattered versus $log(g)$. This is because the targets have different temperature values. Although some targets have nearly the same $log(g)$ values, their temperatures could be dramatically different from each other. In this case, the effects of the variations seen in the right panels can affect the distributions versus the $log(g)$ values. According to our opinion, if we had the targets at the same temperatures, we would obtain a distribution on a linear trend for the two migration periods versus the $log(g)$ values. Indeed, if we compare the targets with the same $log(g)$ values among themselves, as in the left panel of Figure 5, the targets with higher temperatures are located bottom of the panel and the targets with lower temperatures are located at the upper of the panel.

The distribution of spot migration periods is plotted versus the periods of targets in Figure 6. First of all, it is seen from the figure that the spot migration periods decrease with increasing orbital periods. Then, we also see the effect of the stellar temperature on this distribution. In the case of the cool stars, the spot migration periods decrease significantly with increasing orbital periods, while for targets with migration periods less than 800 days, there is no variation observed during their orbital periods.

Finally, to determine the possible reasons for the variations in these frequencies, such as targets' temperatures, age or evolutionary status, the frequencies obtained from the analysis of the ${\theta}_{min}$ residuals and listed in Tables 4-6 were plotted against $B-V$ colour indexes, $log(g)$ values, and orbital period of targets in Figure 7. Examining the frequency distributions in the figure indicates that the frequencies were generally observed for all $B-V$ colour indexes and $log(g)$ values. Although we expected a clear distribution of the frequencies over orbital periods, there is no systematic distribution observed over orbital periods.

\section{Conclusion}\label{S4}
When we considered the general results, it is seen that the stellar spot migration periods were varied as a function of the $B-V$ colour index and exhibited a regular change. It is also recognized that the spot migration periods increase toward the cooler stars in the main sequence. However, a regular variation versus both the parameter $log(g)$ and the rotation periods were also noticed. On the other hand, the effect of temperature is seen as a dominant parameter in these variations. To obtain more conclusive results, the spot migration periods should be examined versus different $log(g)$ values in the case of the targets whose temperatures are the same. By the way, the interpretations for the frequency variations obtained by the ${\theta}_{min}$ analyses are still tentative. More data from stars should be analyzed to obtain a clear conclusion.


\begin{figure*}
\centering
\includegraphics[width=0.8\linewidth]{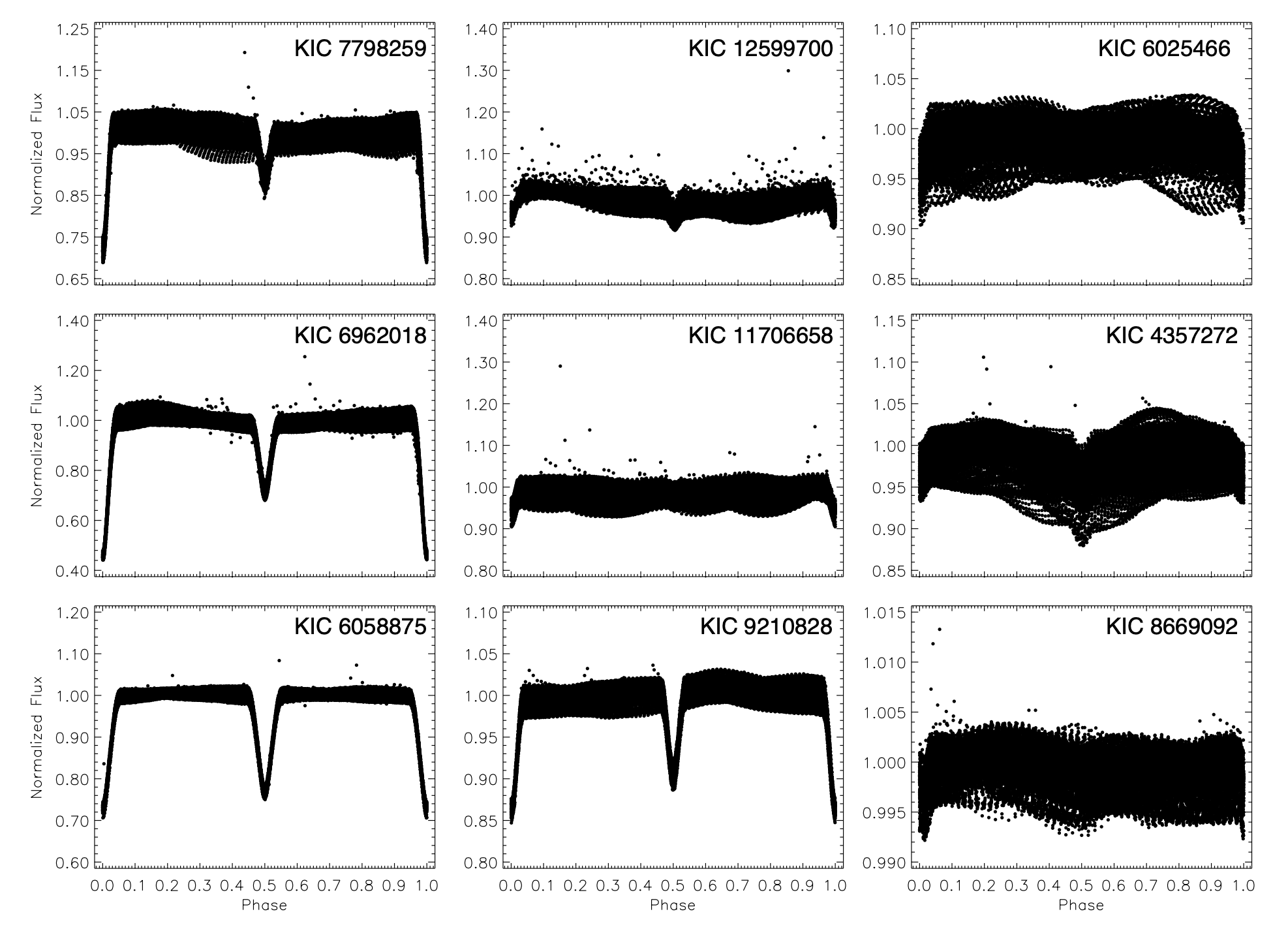}
\caption{Long cadence data of the nine systems from $Kepler$ Database.}
\label{F1}
\end{figure*}

\begin{figure*}
\centering
\includegraphics[width=0.9\linewidth]{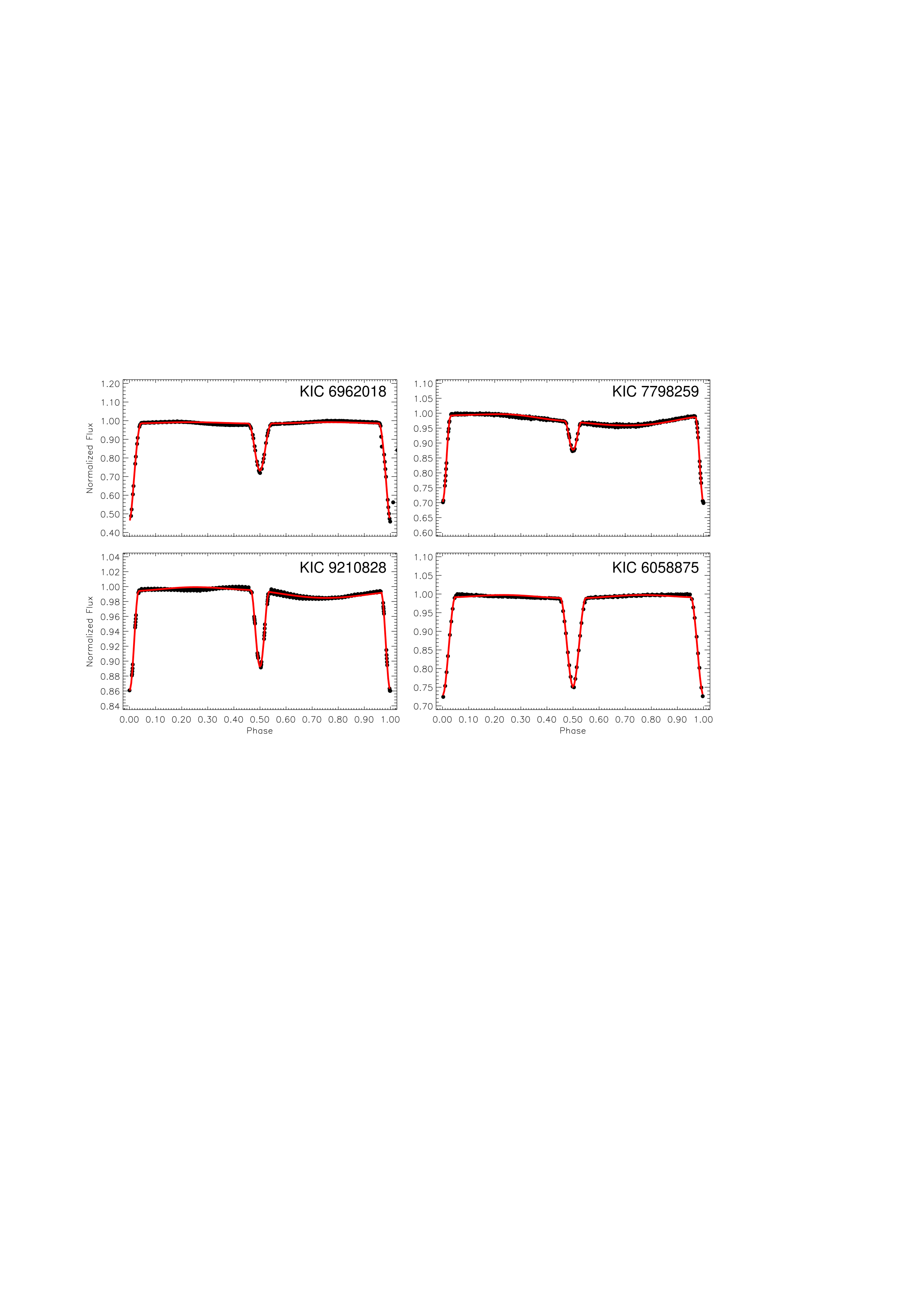}
\caption{The cycles with the least impact of cool stellar spots for KIC 6962018, KIC 7798259, KIC 9210828, and KIC 6058875 systems with their synthetic light curves (red solid line) obtained from the analysis.}
\label{F2}
\end{figure*}

\begin{figure*}
\centering
\includegraphics[width=0.6\linewidth]{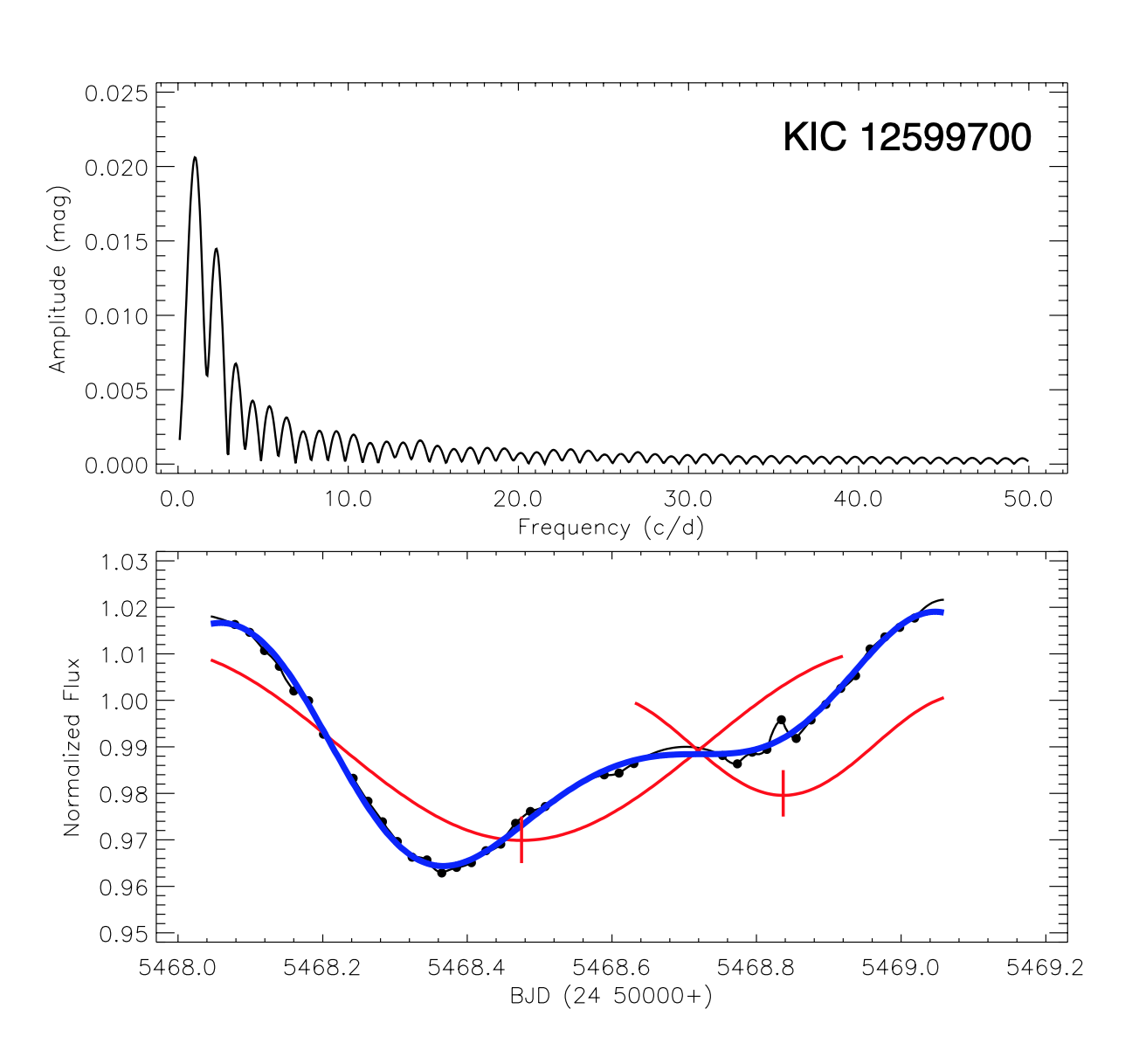}
\caption{The method applied to each cycle in the light curves to determine the minimum point of the spotted area is shown. In the upper panel, the frequencies determined in a single cycle are shown. In the lower panel, an example cycle is shown with its Fourier transform and the minimum times of the cycle. The dots are from the $Kepler$ Eclipsing Binary Archive, representing the long cadence data, while the black straight line is a synthetic curve created to fill in the spaces in observations. The red curves represent the Fourier model corresponding to the two frequencies in the upper panel, while the blue curve is a blended version of these two frequency values. The red vertical lines indicate the minimum points of the curve representing the longitudinal value of the spotted areas.}
\label{F3}
\end{figure*}

\begin{figure*}
\centering
\includegraphics[width=0.6\linewidth]{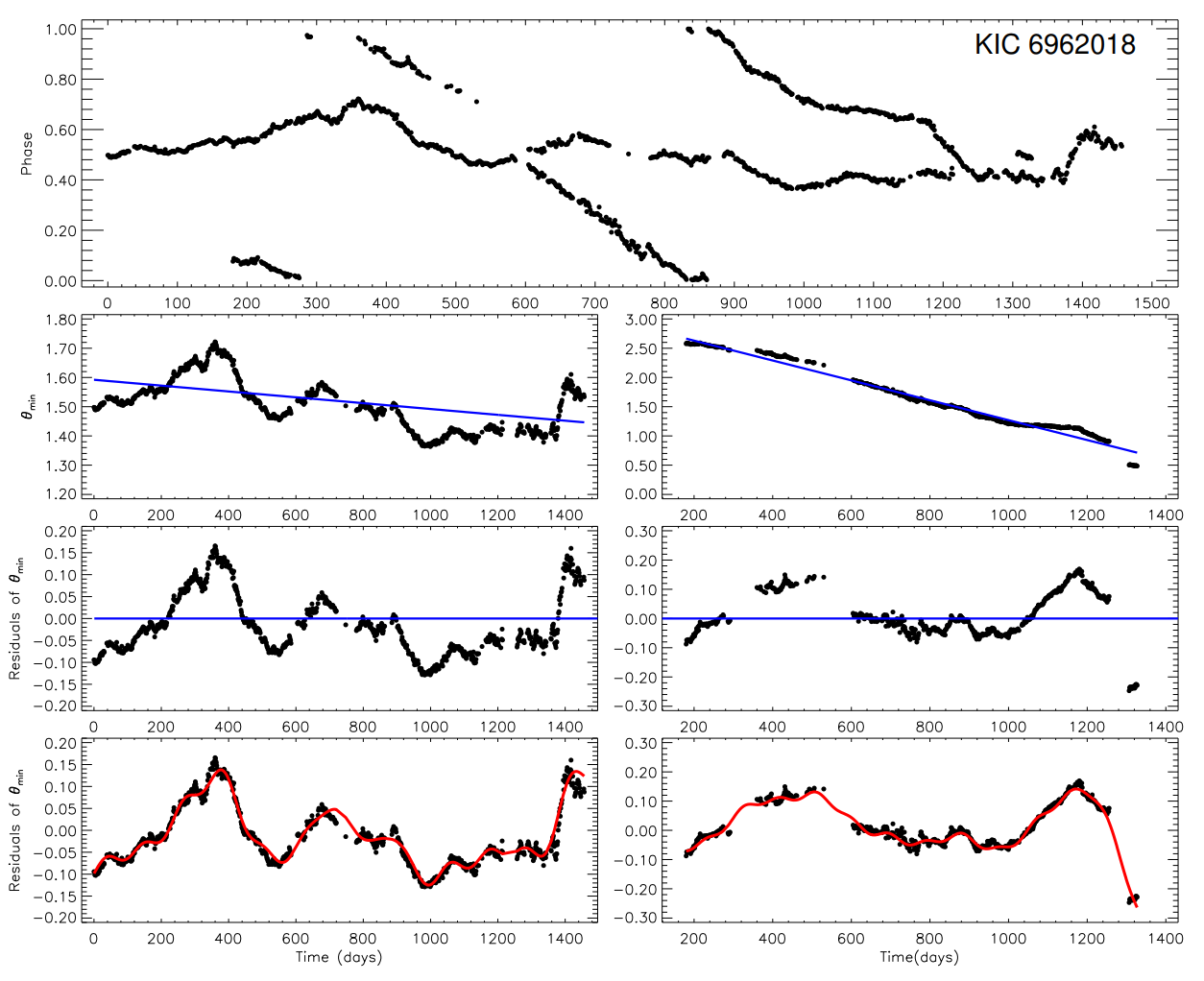}
\caption{The longitudinal migration, the ${\theta}_{min}$ changes and the ${\theta}_{min}$ residuals of cool stellar spots on the active component of KIC 6962018 obtained from the long cadence data. In addition, the representations of the ${\theta}_{min}$ residuals of spots belonging to active components of KIC 6962018, obtained by the frequency analysis are in the bottom panels. The dots show the residuals of ${\theta}_{min}$ and red solid lines represent their representations obtained as a result of the frequency analysis.}
\label{F4}
\end{figure*}

\begin{figure*}
\centering
\includegraphics[width=0.6\linewidth]{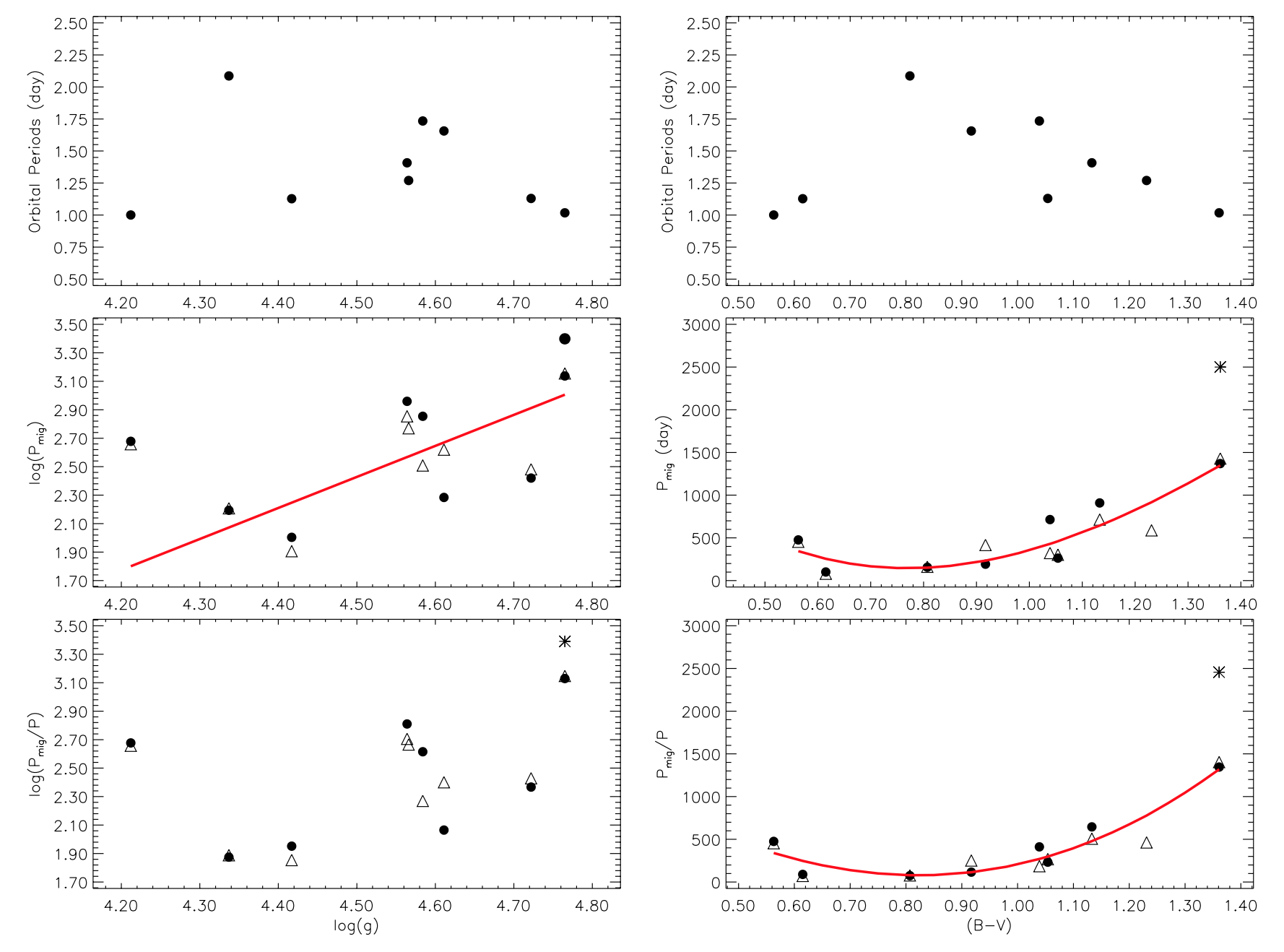}
\caption{The longitudinal migration periods of the spots are plotted against the values of log(g) and $B-V$ colour index of the systems. The dots correspond to longitudinal migration periods of the first spots, the triangles correspond to the second spots, and the star corresponds to the third spot. The red lines in the middle and bottom panels represent the theoretical curves.} 
\label{F5}
\end{figure*}

\begin{figure*}
\centering
\includegraphics[width=0.6\linewidth]{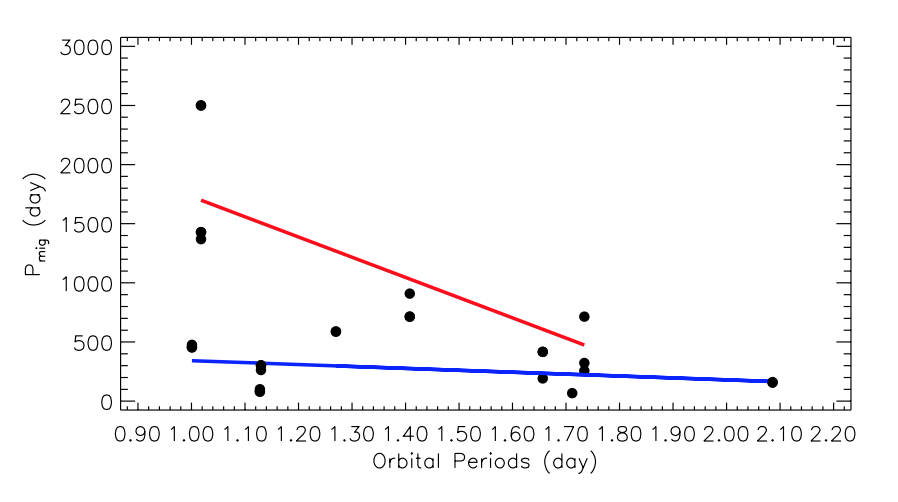}
\caption{The longitudinal migration periods of the spots versus values of periods of the systems. The red solid line represents the change in the migration period of the spots on the cool stars, and the blue solid line represents the change in the migration period of the spots on the hot stars.}
\label{F6}
\end{figure*}

\begin{figure*}
\centering
\includegraphics[width=0.6\linewidth]{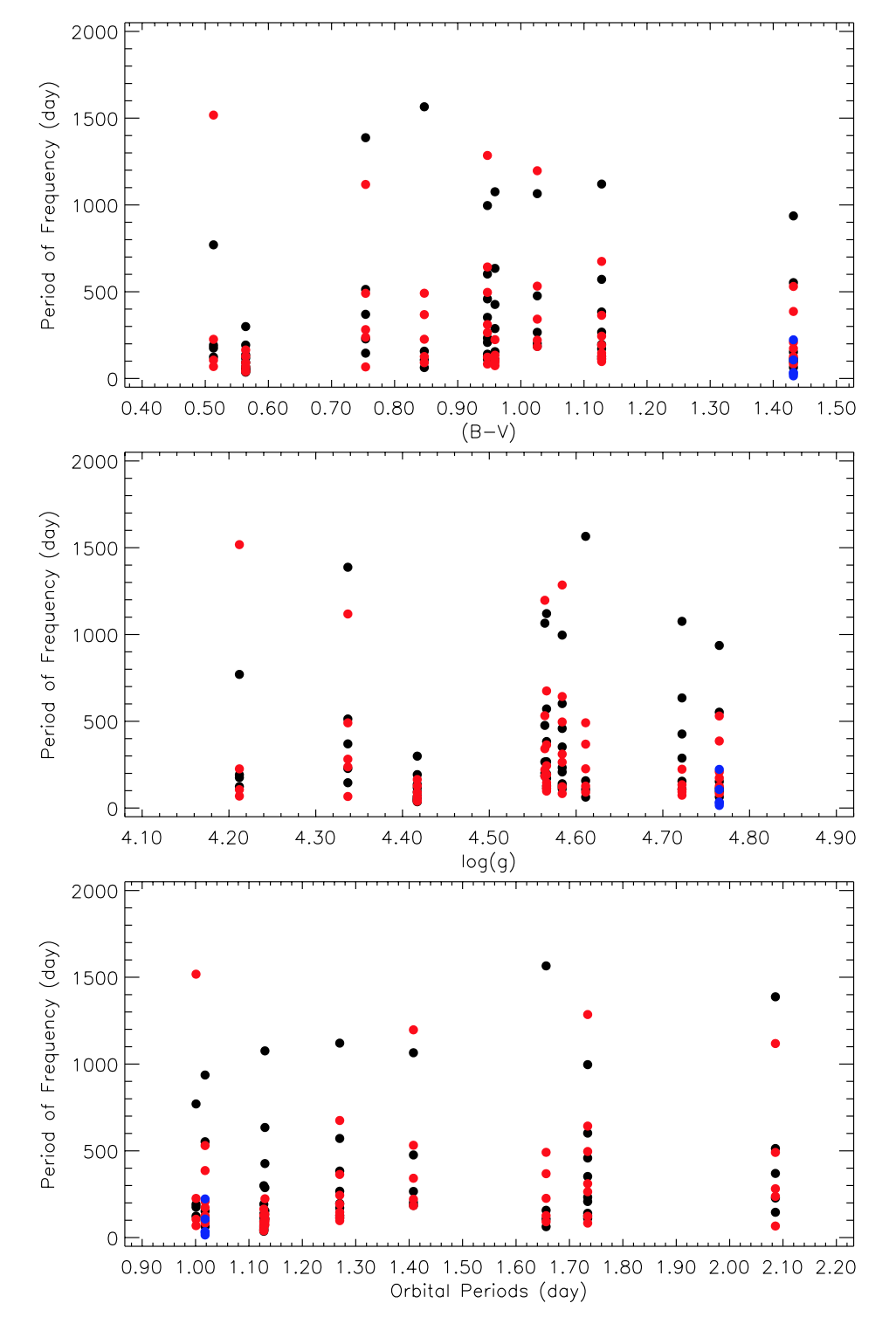}
\caption{The variation of the calculated periods of the frequencies obtained from the residuals of ${\theta}_{min}$and listed in Table 4-6, according to $B-V$, log (g) and period of the systems' components is shown. The dots represent the frequency periods obtained from the first spots, the red dots represent the second spots, and the blue dots represent the third spot.}
\label{F7}
\end{figure*}

\clearpage

\begin{table*}
\begin{threeparttable}
\caption{The physical parameters of systems.}
\label{T1}
\begin{tabular}{cccccccccc}
\toprule
\headrow KIC	&	$T_{0}$$^{1}$ (days)	&	P (days) $^{1}$	&	log(g)$(cm/s^2)$ $^{2}$	& $T_{eff} (K)$ $^{2}$&	$T_{1}$ (K)$^{3}$	&	$T_{2}$ (K)$^{3}$	&	$B-V^{(m)}$	& [Fe/H] $^{4}$	&	i $(^\circ)$ $^{5}$\\ 
\midrule
12599700	&	54965.25916	&	1.01779	&	4.765	&3887 &	3758	&	3759	&	1.361	&	0.17	&	73.23	\\
06962018	&	54965.07384	&	1.26989	&	4.566	&4191 &	3698	&	4539	&	1.231	&	0.36	&	85.44	\\
11706658	&	54965.26679	&	1.40794	&	4.564	&4409 &	4406	&	4343	&	1.133	&	0.16	&	66.33	\\
06058875	&	55002.07258	&	1.12987	&	4.722	&4586 &	5124	&	4855	&	1.054	&	-0.72	&	-	\\
07798259	&	54965.83256	&	1.73422	&	4.584	&4619 &	4952	&	3800	&	1.039	&	-0.19	&	84.29	\\
09210828	&	54965.40883	&	1.65641	&	4.611	&4893 &	5053	&	5236	&	0.917	&	-0.66	&	79.37	\\
04357272	&	54964.96468	&	2.08595	&	4.337	&5117 &	5378	&	4607	&	0.807	&	-0.631	&	80.38	\\
06025466	&	54953.77410	&	1.12780	&	4.417	&5815 &	6367	&	5178	&	0.615	&	-0.22	&	80.64	\\
08669092	&	54954.28886	&	1.00083	&	4.212	&6028 &	6435	&	4785	&	0.563	&	-0.83	&	88.77	\\
\bottomrule
\end{tabular}
\begin{tablenotes}[hang]
\item[]Table note
\item[1]Taken from The Kepler Input Catalogue as $Epoch_{0}$ (BJD 24 00000+).
\item[2]Taken from The Mikulski Archive for Space Telescopes (MAST).
\item[3]\citet{Arm14} 
\item[4]\citet{Pin12} 
\item[5]\citet{Sla11} 
\end{tablenotes}
\end{threeparttable}
\end{table*}

\begin{table*}
\begin{threeparttable}
\caption{The parameters obtained from light curve analysis for 4 binary systems.}
\label{tab2}
\begin{tabular}{@{}lrrrr@{}}
\toprule
\headrow Parameter	&	KIC  06962018	&	KIC 06058875	&	KIC 07798259	&	KIC 09210828	\\
\midrule
$q$	&	0.8878$\pm$0.0307	&	0.9650$\pm$0.0041	&	0.7889$\pm$0.0088	&	0.8965$\pm$0.0066	\\
$i$ $ (^\circ)$	&	87.75$\pm$0.19	&	82.55$\pm$0.06	&	86.27$\pm$0.09	&	80.28$\pm$0.05	\\
$T_{1}$ (K)	&	3698 (Fixed)	&	5124 (Fixed)	&	4952 (Fixed)	&	5053 (Fixed)	\\
$T_{2}$ (K)	&	3421$\pm$60	&	4998$\pm$82	&	4234$\pm$80	&	4795$\pm$50	\\
$\Omega_{1}$	&	7.5036$\pm$0.0730	&	6.5377$\pm$0.0414	&	8.6356$\pm$0.0403	&	7.8340$\pm$0.0345	\\
$\Omega_{2}$	&	8.3470$\pm$0.2528	&	7.9845$\pm$0.0611	&	12.2243$\pm$0.1322	&	8.3118$\pm$0.0538	\\
$L_{1}/L_{T} $	&	0.7430$\pm$0.0154	&	0.6570$\pm$0.0171	&	0.9077$\pm$0.0046	&	0.6483$\pm$0.0093	\\
$g_{1}$, $g_{2}$	&	0.32, 0.32 (Fixed)	&	0.32, 0.32 (Fixed)	&	0.32, 0.32 (Fixed)	&	0.32, 0.32 (Fixed)	\\
$A_{1}$, $A_{2}$	&	0.50, 0.50 (Fixed)	&	0.60, 0.60 (Fixed)	&	0.50, 0.50 (Fixed)	&	0.50, 0.50 (Fixed)	\\
$x_{1},_{bol}$, $x_{2},_{bol}$	&	0.626, 0.626  (Fixed)	&	0.644, 0.644  (Fixed)	&	0.618, 0.618  (Fixed)	&	0.614, 0.614  (Fixed)	\\
$x_{1}$, $x_{2}$	&	0.807, 0.807  (Fixed)	&	0.720, 0.719  (Fixed)	&	0.759, 0.759  (Fixed)	&	0.753, 0.753 (Fixed)	\\
$< r_{1} >$$^{a}$	&	0.1513$\pm$0.0018	&	0.1805$\pm$0.0013	&	0.1281$\pm$0.0006	&	0.1450$\pm$0.0007	\\
$< r_{2} >$$^{b}$	&	0.1224$\pm$0.0041	&	0.1389$\pm$0.0012	&	0.0695$\pm$0.0008	&	0.1241$\pm$0.0009	\\
\bottomrule
\end{tabular}
\begin{tablenotes}[hang]
\item[]Table note
\item[a,b] $<r_{1}>$ and $<r_{2}>$ are the fractional radii (real radius / semi-major axis, $R/a$) for the primary and secondary components, respectively.
\end{tablenotes}
\end{threeparttable}
\end{table*}

\begin{table*}
\begin{threeparttable}
\caption{The longitudinal migration periods of cool stellar spots on active components of systems.}
\label{tab3}
\begin{tabular}{@{}lccc@{}}
\toprule
\headrow Name (KIC) & $P_{Mig 1}$ (days) & $P_{Mig 2}$ (days) & $P_{Mig 3}$ (days) \\
\midrule
12599700	&	2500.000$\pm$0.004	&	1428.571$\pm$0.082	&	1369.863$\pm$0.027	\\
06962018	&	-	&	588.235$\pm$0.022		&	-	\\
11706658	&	909.091$\pm$0.036		&	714.286$\pm$0.111		&	-	\\
06058875	&	263.158$\pm$0.101		&	303.030$\pm$0.329		&	-	\\
07798259	&	714.286$\pm$0.151		&	322.581$\pm$0.846		&	-	\\
09210828	&	192.308$\pm$0.135		&	416.667$\pm$0.078	    &	-	\\
04357272	&	156.250$\pm$0.070		&	161.290$\pm$0.064		&	-	\\
06025466	&	101.010$\pm$0.057		&	80.645$\pm$0.086		&	-	\\
08669092	&	476.190$\pm$0.062		&	454.545$\pm$0.555		&	-	\\
\bottomrule
\end{tabular}
\end{threeparttable}
\end{table*}

\clearpage

\begin{table*}
\begin{threeparttable}
\caption{The values of frequency and period obtained as a result of frequency analysis of ${\theta}_{min}$ residuals.}
\label{T4}
\begin{tabular}{llcrrlcrr}
\toprule
\headrow Name	& \multicolumn{4}{c}{First Active Region}   &\multicolumn{4}{c}{Second Active Region}  \\
\headrow(KIC)	&	 No. 	&	 Frequency ($d^{-1}$)	&	 Period (d) 	&SNR & No. &	 Frequency ($d^{-1}$)	&	 Period (d) 	&	SNR	 \\	
\midrule
12599700	&	F1	&	0.00181	$\pm$	0.00002	&	552.511	&	16.634	&	F1	&	0.00259	$\pm$	0.00000	&	386.189	&	27.231	\\
	&	F2	&	0.00654	$\pm$	0.00002	&	152.822	&	11.594	&	F2	&	0.00189	$\pm$	0.00000	&	530.449	&	18.315	\\
	&	F3	&	0.00877	$\pm$	0.00004	&	114.010	&	6.211	&	F3	&	0.00479	$\pm$	0.00001	&	208.713	&	9.292	\\
	&	F4	&	0.00107	$\pm$	0.00004	&	936.866	&	7.042	&	F4	&	0.00577	$\pm$	0.00001	&	173.263	&	9.633	\\
	&	F5	&	0.01564	$\pm$	0.00005	&	63.940	&	5.653	&	F5	&	0.00793	$\pm$	0.00001	&	126.028	&	6.659	\\
	&	F6	&	0.01035	$\pm$	0.00005	&	96.627	&	5.084	&	F6	&	0.01194	$\pm$	0.00002	&	83.770	&	3.683	\\
\midrule
6962018	&	F1	&	0.00089	$\pm$	0.00000	&	1120.472	&	27.082	&	F1	&	0.00148	$\pm$	0.00001	&	674.843	&	15.137	\\
	&	F2	&	0.00175	$\pm$	0.00001	&	571.221	&	21.002	&	F2	&	0.00275	$\pm$	0.00001	&	364.201	&	15.634	\\
	&	F3	&	0.00261	$\pm$	0.00001	&	383.319	&	18.634	&	F3	&	0.00410	$\pm$	0.00001	&	244.092	&	13.258	\\
	&	F4	&	0.00374	$\pm$	0.00001	&	267.269	&	10.961	&	F4	&	0.00519	$\pm$	0.00001	&	192.812	&	9.783	\\
	&	F5	&	0.00783	$\pm$	0.00003	&	127.773	&	4.368	&	F5	&	0.00793	$\pm$	0.00002	&	126.070	&	8.096	\\
	&	F6	&	0.00584	$\pm$	0.00002	&	171.366	&	5.307	&	F6	&	0.00684	$\pm$	0.00003	&	146.144	&	5.221	\\
	&	F7	&	0.00879	$\pm$	0.00003	&	113.798	&	4.239	&	F7	&	0.00893	$\pm$	0.00002	&	111.925	&	5.578	\\
	&	F8	&	0.00508	$\pm$	0.00003	&	196.840	&	3.759	&	F8	&	0.01024	$\pm$	0.00003	&	97.637	&	4.165	\\

\midrule
11706658	&	F1	&	0.00094	$\pm$	0.00000	&	1065.148	&	32.533	&	F1	&	0.00084	$\pm$	0.00000	&	1197.112	&	32.973	\\
	&	F2	&	0.00210	$\pm$	0.00000	&	476.216	&	17.908	&	F2	&	0.00188	$\pm$	0.00000	&	532.050	&	15.376	\\
	&	F3	&	0.00376	$\pm$	0.00001	&	266.226	&	6.727	&	F3	&	0.00292	$\pm$	0.00001	&	342.032	&	7.598	\\
	&	F4	&	0.00496	$\pm$	0.00001	&	201.752	&	5.113	&	F4	&	0.00452	$\pm$	0.00001	&	221.005	&	4.578	\\
	&	F5	&	0.00539	$\pm$	0.00001	&	185.366	&	4.761	&	F5	&	0.00042	$\pm$	0.00001	&	2394.224	&	3.936	\\

\midrule
6058875	&	F1	&	0.00093	$\pm$	0.00000	&	1075.889	&	31.051	&	F1	&	0.00745	$\pm$	0.00001	&	134.231	&	10.231	\\
	&	F2	&	0.00234	$\pm$	0.00001	&	426.646	&	12.596	&	F2	&	0.01076	$\pm$	0.00002	&	92.936	&	5.887	\\
	&	F3	&	0.00348	$\pm$	0.00001	&	287.738	&	8.221	&	F3	&	0.01352	$\pm$	0.00002	&	73.946	&	7.599	\\
	&	F4	&	0.00158	$\pm$	0.00001	&	634.499	&	5.991	&	F4	&	0.00447	$\pm$	0.00002	&	223.750	&	5.982	\\
	&	F5	&	0.00917	$\pm$	0.00002	&	109.011	&	3.910	&	F5	&	0.00933	$\pm$	0.00003	&	107.157	&	3.571	\\
	&	F6	&	0.00647	$\pm$	0.00002	&	154.659	&	3.748	&		&		-		&		-&	-	\\
\midrule
7798259	&	F1	&	0.00166	$\pm$	0.00000	&	602.079	&	26.423	&	F1	&	0.00156	$\pm$	0.00000	&	642.612	&	20.895	\\
	&	F2	&	0.00218	$\pm$	0.00001	&	458.727	&	23.758	&	F2	&	0.00078	$\pm$	0.00001	&	1285.224	&	9.155	\\
	&	F3	&	0.00429	$\pm$	0.00002	&	233.063	&	7.394	&	F3	&	0.00813	$\pm$	0.00001	&	122.934	&	7.848	\\
	&	F4	&	0.00100	$\pm$	0.00001	&	996.544	&	11.079	&	F4	&	0.00322	$\pm$	0.00001	&	310.713	&	7.192	\\
	&	F5	&	0.00713	$\pm$	0.00002	&	140.290	&	6.367	&	F5	&	0.01199	$\pm$	0.00002	&	83.407	&	3.365	\\
	&	F6	&	0.00903	$\pm$	0.00003	&	110.727	&	4.585	&		&		-		&	-	&	-	\\
	&	F7	&	0.00284	$\pm$	0.00002	&	352.436	&	5.579	&		&		-		&	-	&	-	\\
	&	F8	&	0.00481	$\pm$	0.00002	&	207.912	&	5.007	&		&		-		&	-	&	-	\\
\bottomrule
\end{tabular}
\end{threeparttable}
\end{table*}

\clearpage

\begin{table*}
\begin{threeparttable}
\caption{Continued.}
\label{T5}
\begin{tabular}{llcrrlcrr}
\toprule
\headrow Name	& \multicolumn{4}{c}{First Active Region}   &\multicolumn{4}{c}{Second Active Region}  \\
\headrow(KIC)	&	 No. 	&	 Frequency ($d^{-1}$)	&	 Period (d) 	&SNR & No. &	 Frequency ($d^{-1}$)	&	 Period (d) 	&	SNR	 \\		
\midrule
9210828	&	F1	&	0.00635	$\pm$	0.00001	&	157.427	&	11.045	&	F1	&	0.00204	$\pm$	0.00001	&	491.281	&	21.324	\\
	&	F2	&	0.00064	$\pm$	0.00002	&	1565.525	&	7.673	&	F2	&	0.00272	$\pm$	0.00001	&	368.312	&	18.516	\\
	&	F3	&	0.00930	$\pm$	0.00002	&	107.555	&	8.332	&	F3	&	0.00442	$\pm$	0.00002	&	226.253	&	10.810	\\
	&	F4	&	0.01586	$\pm$	0.00003	&	63.041	&	3.775	&	F4	&	0.00790	$\pm$	0.00003	&	126.588	&	6.355	\\

\midrule
4357272	&	F1	&	0.00072	$\pm$	0.00000	&	1387.578	&	20.502	&	F1	&	0.00089	$\pm$	0.00001	&	1118.274	&	15.736	\\
	&	F2	&	0.00270	$\pm$	0.00001	&	369.781	&	6.485	&	F2	&	0.00204	$\pm$	0.00001	&	490.749	&	7.498	\\
	&	F3	&	0.00685	$\pm$	0.00001	&	145.929	&	5.169	&	F3	&	0.00355	$\pm$	0.00002	&	281.586	&	4.083	\\
	&	F4	&	0.00195	$\pm$	0.00002	&	513.376	&	3.224	&		&		-		&	-	&	-	\\
	&	F5	&	0.00439	$\pm$	0.00002	&	228.042	&	3.818	&		&		-		&	-	&	-	\\

\midrule
6025466	&	F1	&	0.00334	$\pm$	0.00002	&	299.201	&	9.316	&	F1	&	0.00783	$\pm$	0.00001	&	127.715	&	9.725	\\
	&	F2	&	0.01682	$\pm$	0.00002	&	59.469	&	10.864	&	F2	&	0.02536	$\pm$	0.00001	&	39.425	&	7.622	\\
	&	F3	&	0.00735	$\pm$	0.00002	&	136.129	&	9.710	&	F3	&	0.01516	$\pm$	0.00002	&	65.962	&	5.628	\\
	&	F4	&	0.00519	$\pm$	0.00003	&	192.774	&	8.222	&	F4	&	0.00608	$\pm$	0.00001	&	164.408	&	8.584	\\
	&	F5	&	0.01904	$\pm$	0.00002	&	52.511	&	12.047	&	F5	&	0.01103	$\pm$	0.00003	&	90.647	&	4.200	\\
	&	F6	&	0.02141	$\pm$	0.00003	&	46.705	&	7.422	&	F6	&	0.02217	$\pm$	0.00003	&	45.111	&	3.614	\\
	&	F7	&	0.01079	$\pm$	0.00003	&	92.656	&	7.744	&		&		-		&	-	&	-	\\
	&	F8	&	0.02740	$\pm$	0.00004	&	36.497	&	5.496	&		&		-		&	-	&	-	\\
	&	F9	&	0.00853	$\pm$	0.00003	&	117.238	&	6.745	&		&		-		&	-	&	-	\\
	&	F10	&	0.01469	$\pm$	0.00004	&	68.065	&	5.311	&		&		-		&	-	&	-	\\
	&	F11	&	0.02353	$\pm$	0.00005	&	42.490	&	4.440	&		&		-		&	-	&	-	\\

\midrule
8669092	&	F1	&	0.00522	$\pm$	0.00004	&	191.655	&	8.520	&	F1	&	0.00066	$\pm$	0.00005	&	1517.752	&	7.058	\\
	&	F2	&	0.00570	$\pm$	0.00005	&	175.528	&	5.889	&	F2	&	0.00443	$\pm$	0.00005	&	225.867	&	7.342	\\
	&	F3	&	0.00808	$\pm$	0.00010	&	123.737	&	3.308	&		&		-		&	-	&	-	\\

\bottomrule
\end{tabular}
\end{threeparttable}
\end{table*}

\begin{table*}
\begin{threeparttable}
\caption{Continued.}
\label{T6}
\begin{tabular}{llcrr}
\toprule
\headrow Name	& \multicolumn{4}{c}{Third Active Region}   \\
\headrow(KIC)	&	 No. 	&	 Frequency ($d^{-1}$)	&	 Period (d) 	&SNR \\
\midrule
12599700	&	F1	&	0.00935	$\pm$	0.00015	&	106.980	&	7.806	\\
&F2	&	0.00449	$\pm$	0.00019	&	222.876	&	6.028	\\
&F3	&	0.03440	$\pm$	0.00032	&	29.071	&	3.776	\\
&F4	&	0.02898	$\pm$	0.00040	&	34.510	&	3.039	\\
\bottomrule
\end{tabular}
\end{threeparttable}
\end{table*}

\clearpage
\printbibliography

\end{document}